\documentclass[12pt,preprint]{aastex}
\usepackage{epsfig}

\shortauthors{Xue \& Cui} \shorttitle{X-ray Flares from Mrk~501}

\begin{document}

\title{X-RAY FLARES FROM MARKARIAN~501}

\author{Yongquan Xue and Wei Cui}
\affil{Department of Physics, Purdue University, West Lafayette, IN 47907; 
xuey@physics.purdue.edu, cui@physics.purdue.edu}

\begin{abstract}

Motivated by the recent finding of hierarchical X-ray flaring phenomenon 
in Mrk 421, we conducted a systematic search for X-ray flares from Mrk 501, 
another well-known TeV blazar, by making use of the rich {\em RXTE} 
archival database. We detected flares over a wide range of timescales, 
from months down to minutes, as in the case of Mrk 421. However, the flares 
do not seem to occur nearly as frequently in Mrk 501 as in Mrk 421 on any 
of the timescales. The flaring hierarchy also seems apparent in Mrk 501, 
suggesting that it might be common among TeV blazars. The results seem to 
imply a scale-invariant physical origin of the flares (large or small). 
The X-ray spectrum of the source shows a general trend of hardening toward 
the peak of long-duration flares, with indication of spectral hysteresis, 
which is often seen in TeV blazars. However, the data are not of sufficient 
quality to allow us to draw definitive conclusions about spectral variability 
associated with more rapid but weaker flares. We critically examine a 
reported sub-hour X-ray flare from Mrk 501, in light of intense background 
flaring activity at the time of the observation, and concluded that the 
flare is likely an artifact. On the other hand, we did identify a rapid 
X-ray flare that appears to be real. It lasted only for about 15 minutes, 
during which the flux of the source varied by about 30\%. Sub-structures 
are apparent in its profile, implying variabilities on even shorter 
timescales. Such rapid variabilities of Mrk 501 place severe constraints 
on the physical properties of the flaring region in the jet, which have 
serious implications on the emission models proposed for TeV blazars.

\end{abstract}


\section{Introduction}

Mrk 501 is among the brightest and closest ($z=0.034$) extragalactic X-ray 
sources in the sky. It is classified as a BL Lacertae object and is located 
in the elliptical galaxy UGC 10599 (Stickel et al. 1993). BL Lac objects 
belong to a more general class of radio-loud active galactic nuclei known 
as blazars, which are characterized by rapid variability and non-thermal 
emission at nearly all wavelengths. The emission from a blazar is generally
thought to be dominated by radiation from a relativistic jet that is 
directed roughly along the line of sight (Urry \& Padovani 1995). The 
spectral energy distribution (SED) of blazars invariably shows two 
characteristic ``humps'' in the $\nu F_{\nu}$ representation, with one 
located at optical--X-ray energies and the other at GeV--TeV energies 
(Fossati et al. 1998). Mrk 501 is among a small number of blazars that 
have been detected at TeV energies (e.g., Quinn et al. 1996; Bradbury et 
al. 1997).

For TeV blazars, there is a general correlation between fluxes at X-ray 
and TeV energies (where the SED peaks), although exceptions have recently
been noted (Krawczynski et al. 2004; Cui et al. 2004). While the origin of 
gamma rays is still being debated, there is a general consensus that 
X-rays originate in the synchrotron radiation from highly relativistic 
electrons in the jet. Two classes of models have been proposed to explain
gamma-ray emission from blazars. In the leptonic models, the TeV emission 
is attributed to the synchrotron self-Compton (SSC) process (e.g., Marscher 
\& Gear 1985; Maraschi et al. 1992). The SSC models are attractive for 
their conceptual simplicity. They can quite naturally account for the 
observed X-ray--TeV correlation, and have also enjoyed some success in 
fitting the observed SEDs (e.g., Kataoka et al. 1999; Sambruna et al. 2000; 
Petry et al. 2000; Krawczynski et al. 2002; Konopelko et al. 2003). However, 
such models still face challenges, such as the presence of ``orphan TeV 
flares'' (i.e., those TeV flares with no apparent counterparts in X-rays; 
Krawczynski et al. 2004; Cui et al. 2004). In the hadronic models, the TeV 
emission is probably due to synchrotron radiation from relativistic protons 
in the jet (Aharonian 2000; M\"ucke et al. 2003), although other hadronic
processes might also contribute. Although the models are being pushed to 
the limit by the observed rapid variability of TeV blzars, they are, by no 
means, ruled out yet. It remains to be seen whether they can account for 
the X-ray--TeV correlation in a quantitative manner. 

TeV blazars are known to undergo flaring activities both at X-ray and TeV 
energies. The flares seem to occur on all timescales but vary greatly in 
magnitude, as illustrated vividly by the X-ray flaring hierarchy in Mrk 421 
(Cui 2004). The hierarchy strongly implies the scale-invariant nature of 
flaring processes in the source, prompting comparison to solar flares and 
seemingly similar rapid flares in stellar-mass black hole systems. The 
flares from Mrk 501 have been well observed at X-ray and TeV energies (e.g., 
Quinn et al. 1996; Bradbury et al. 1997; Catanese et al. 1997; Hayashida 
et al. 1998; Kataoka et al. 1999; Sambruna et al. 2000; Petry et al. 2000).
In fact, the detection of a very rapid X-ray flare from the source was 
reported (Catanese \& Sambruna 2000), which represented the first case of 
sub-hour X-ray flares in any TeV blazar. The short duration of the flare 
would lead to severe constraints on the properties of the flaring region, 
as in the case of Mrk 421 (Cui 2004).

Motivated by the recent finding of hierarchical X-ray flaring phenomenon 
in Mrk 421 (Cui 2004), we conducted a systematic search for X-ray flares 
in Mrk 501 over a broad range of timescales, making use of the rich 
{\em RXTE} archival database. We describe the data and data reduction 
procedure in \S~2. The results are presented in \S~3, along with a 
critical examination of the reported sub-hour X-ray flare. Finally, 
in \S~4, we discuss the results and their implications on the proposed 
models.

\section{Data and Data Reduction}

Mrk 501 has been observed frequently with {\em RXTE}. For this work, we
obtained publicly available data from the archival databases,
including data both from the All-Sky Monitor (ASM) and the Proportional 
Counter Array (PCA). We obtained the ASM light curves of Mrk~501 from the 
MIT archive.\footnote{See http://xte.mit.edu/asmlc/srcs/mkn501.html\#data} 
We chose to filter out data points with error bars (on the raw count rates 
in the summed band) greater than 2.0 counts/s, which constitute about 5\% 
of the data. We then weighted the rates by $1/\sigma^2$ and rebinned 
them to produce light curves with 21-day time bins. The light curves are 
available in three energy bands: 1.5--3, 3--5, and 5--12 keV, providing 
crude spectral information. 

The PCA is a narrow-field, pointing instrument. It consists of five nearly 
identical proportional counter units (PCUs) and covers a nominal energy 
range of 2--60 keV. Due to operational
constraints, some of the PCUs were often switched off. Only PCU 0 and PCU 2 
are nearly always in operation. Data analysis is further complicated by 
the loss of the front veto layer in PCU 0 (in 2000), because the data from 
PCU 0 is now more prone to contamination by events caused by low-energy 
electrons entering the detector. This is particularly relevant to the study
of variability of weak sources like Mrk~501. Fortunately, it only affects a 
about 15\% of the data used in this work. We excluded from timing analyses 
PCU 0 data collected in 2000. Most PCA observations were made in ``snap-shot'' 
modes, with typical effective exposure times ranging from half a kilosecond 
to a few kiloseconds. The PCA has numerous data modes and multiple modes 
are usually employed in an observation. For this work, however, we only 
used the {\em Standard 2} data, which have a time resolution of 16 s. 

We followed Cui (2004) closely in reducing and analyzing the PCA data. 
Briefly, the data were reduced with {\em FTOOLS 5.2}. For an observation, 
we first filtered data by following the standard procedure for faint 
sources (see the online {\em RXTE} Cook Book),\footnote{See
http://heasarc.gsfc.nasa.gov/docs/xte/recipes/cook\_book.html.} which 
resulted in a list of good time intervals (GTIs). We then simulated 
background events for the observation with the latest background model 
that is appropriate for faint sources (pca\_bkgd\_cmfaintl7\_eMv20031123.mdl). 
Using the GTIs, we proceeded to extract a light curve from the data (by 
combining all active PCUs) in each of the following energy bands: 2.0--5.7, 
5.7--11, 11--60, and 2.0--60 keV. Note that the boundaries of each band are 
matched up as closely as possible across different PCA epochs but are only 
approximate (up to $\pm$0.2 keV). We repeated the steps to construct the 
corresponding background light curves from the simulated 
events. Finally, we obtained light curves of the source by subtracting 
off the background. Following a similar procedure, we constructed an X-ray 
spectrum and the associated background spectrum for each observation. We 
note that for spectral analysis we only used data from the first xenon layer 
of each PCU (which is most accurately calibrated). For the purpose of 
detailed spectral modeling (see \S~3.4), we included a 1\% systematic 
uncertainty uniformly across the entire energy range of interest, to 
take into account residual calibration uncertainties. Note that we did 
include PCU 0 data from the 2000 observations in the spectral analysis 
because the background model for the detector seems to work fine in all 
cases.

\section{Results}

\subsection{Long X-ray Flares: Months to $>$ Year}

Figure 1 shows the ASM light curve of Mrk 501 for the 1.5--12 keV band over 
roughly an eight-year period. The source was very active in X-rays in 1997.
The activity continued, at a much lower level, in 1998 and also the early 
part of 1999. Mrk 501 has been relatively quiet ever since. At least two 
major flares are easily identifiable from the light curve, 
which lasted for months to over a year. The flares show substantial 
sub-structures, indicating variability on shorter timescales, perhaps in 
the form of unresolved weaker and shorter flares.

Though crude, hardness ratios make it possible to study, in a 
model-independent manner, spectral variability of a source. Here, we 
computed the ratio between the ASM count rate in the 3--12 keV band to 
that in the 1.5--3 keV band to examine spectral evolution of Mrk 501 
during the 1997 giant flare. The results are shown in Figure 2. For 
clarity, we have averaged the raw ASM count rates with variable time bins 
(6-week, 2-week, and 1-week), depending on the brightness of the source. 
The figure clearly shows a hardening trend of the X-ray spectrum as the 
source approaches the peak of the flare. This is consistent with
findings of previous works (e.g., Pian et al. 1998). Such a 
hardness-intensity correlation seems to hold in general for X-ray bright 
blazars (Giommi et al. 1990).

To investigate the phenomenon of spectral hysteresis, which is often 
observed in blazars (e.g., Takahashi et al. 1996, Kataoka et al. 2000, 
Zhang 2002, Giebels et al. 2002, Falcone et al. 2004, Cui 2004)
but may be too subtle to be seen in the ASM data, we made a similar 
hardness-intensity diagram from the PCA data. There are three observing 
periods, with two covering the rising portion of the 1997 flare and one the 
decaying portion, as shown in Figure~3 (top panel). Unfortunately, the 
coverage of the flare with the PCA is quite sparse, especially during the 
decaying phase. While there appears to be indication of spectral 
hysteresis during the rising phase (perhaps associated with smaller 
flares) and between the rising and decaying phases, it is difficult to 
draw any definitive conclusions from these results. On the other hand, 
the trend of spectral hardening with increasing fluxes is also apparent 
in the PCA data.

\subsection{Intermediate X-ray Flares: Days to Weeks}

The PCA data span a period of 4 years, which covers weaker flaring activities 
in 1998, 1999, and 2000, besides the giant flare in 1997 (see Fig.~1). The 
light curves reveal the presence of X-ray flares with shorter durations, from 
days to weeks, nearly all the time, as shown in Figure~4. Although the ASM 
light curve shows that the source seems to be very quiet during the PCA 
observations in 1999 and 2000, the more sensitive PCA light curves reveal 
flaring activities during these time periods. Therefore, like Mrk 421, there 
is no apparent ``quiescent'' state for Mrk 501 either. Unlike Mrk 421, 
however, the flux of Mrk 501 does not seem to vanish, implying the presence 
of steady-state emission. The amplitude of these intermediate flares can also
be very large, up to 4--5 times the steady-state flux. We attempted to 
investigate spectral evolution and hysteresis of the source across individual 
flares but found that the results were ambiguous, mainly due to the lack of 
quality of the data in this case.

\subsection{Rapid X-ray Flares: $<$ Hour}

Catanese \& Sambruna (2000) reported the first case of a sub-hour X-ray 
flare from Mrk 501. Surprisingly, however, we initially failed to find it in 
the PCA light curves constructed from the same observation. Examining
the light curves more carefully, we realized that there was actually a 
data gap almost exactly at the time of the flare. We suspected that the 
data had probably been filtered out during the data cleaning process in our
case. We then relaxed the data filtering criteria one at a time and in 
the end found that it was the high ``ELECTRON2'' values that deemed the 
time interval (of the reported flare) ``bad''. In general, the ``ELECTRON'' 
parameter provides a measure of contamination of the data by events 
induced by electrons entering the detector. Some of such events (due to 
low energy electrons) may be registered as good X-ray events because 
they fail to trigger any of the vetoing logics. Since the PCU 2 is always 
on, we used ``ELECTRON2'' in the data cleaning procedure (as recommended
in the {\em RXTE} Cookbook).

To show possible electron contamination of the data, we plotted the X-ray
count rates and the values of the ``ELECTRON2'' parameter for the 
observation in Figure~5 (left panels). The figure clearly shows that there 
is a strong electron flare nearly at the same time as the reported X-ray
flare, which raises the possibility of the latter being an artifact. 
However, we note that the X-ray flare seems to have started about 200 s 
before the electron flare and that the profile of the X-ray flare appears 
asymmetric, unlike that of the electron flare. We cannot think of an 
obvious explanation for either ``anomaly''. On the other hand, we found 
another rapid X-ray flare (which has not been reported) from a different 
observation (obsid: 30249-01-01-01), which also has an 
asymmetric profile, is accompanied by an electron flare, and leads the 
electron flare by about the same amount of time ($\sim$200 s), as also 
shown in Fig.~5 (right panels) for a direct comparison. The presence of 
two such flares raised serious doubts in our mind about them being 
physically associated with Mrk 501.

We then systematically searched for X-ray flares that were accompanied by 
electron flares, by visually inspecting all of the light curves, and found 
a total of 21 cases (although it is entirely possilble than some weaker 
ones might have escaped our attention), including the two just discussed. 
Interestingly, in all other 19 cases, we found no apparent timing offsets 
between X-ray and electron flares. Figure~6 shows some representative 
examples of those. It is clear, from the figure, that not all electron 
flares are registered as X-ray signals and that the profile of an electron 
flare can be complex, e.g., with multiple peaks. Moreover, we found that 
the spectrum of the X-ray flares varies greatly. In extreme cases, a single 
X-ray flare is seen in one energy band that coincides with one of the peaks 
of an electron flare, but another flare in a different energy band that 
coincides with a different peak of the same electron flare. Therefore, the 
phenomenology is complex. 

Finally, we derived the latitude and longitude of the {\em RXTE} satellite 
at the time of the peak of each of the 21 electron flares. Figure~7 shows 
a map that summarizes the results. The two cases shown in Fig.~5 are 
highlighted with different symbols. It should be noted that the peaks of 
these two electrons flares are separated by almost exactly one satellite 
orbit ($\approx$1.6 hours). In general, the electron flares seem to cluster 
in the northeast Pacific region, which confirms what is generally known 
about the occurrence of such events (K.~Jahoda and C.~Markwardt of the PCA 
team, private communication). However, particle fluxes in the region are 
usually over two orders of magnitude smaller than those in the well-known 
South Atlantic Anomaly (SAA) region, which is why the region is not 
screened in the same manner as the SAA. Putting 
together all the evidences, we think that the reported rapid flare (and 
also the one found in this work) is most likely an artifact caused by 
intense flaring in soft electron fluxes. Admittedly, outstanding issues 
remain regarding the observed timing offsets or asymmetric profiles (see 
Fig.~5). Neither can we rule out the possibility that real flares from 
Mrk 501 might sometimes include a very soft second delayed component, 
which increases the ``ELECTRON'' parameter through extra counts absorbed 
in the propane layer, although the scenario seems a bit contrived.

In our search, we discovered yet another sub-hour X-ray flare. It occurred 
during an observation (obsid: 30249-01-01-02) when there was hardly any 
enhancement in the soft electron fluxes. Figure~8 shows the flare, along 
with the ``ELECTRON2'' parameter for the same time period. The flare is 
quite weak (compared with the other two), reaching a peak amplitude of 
about 30\% of the steady-state flux. It lasted 
for about 800 s and shows sub-structures on even shorter timescales. The 
profile shown might represent the composite of two flares or a sudden 
decrease and recovery of X-ray fluxes in the middle of a flare. Remarkably, 
the rising or decaying time associated with the sub-structures is as short 
as about 20--30 seconds! The flare occurred roughly one satellite orbit
after the reported one, which might be a bit worrisome. However, it would 
appear about 50\arcdeg\ in longitude further west (as shown in Fig.~7). 
Moreover, Mrk 501 was monitored very intensely during this time period, 
with data taken in 18 nearly consecutive satellite orbits. Most of these 
observations did indeed span periods when the satellite was in the general 
area of the ``Northeast Pacific Region'', but no X-ray/electron flare 
(other than the ones already mentioned) was detected, which rules out
any systematic effects associated with the region. Finally, if the flare 
shown in Fig. 8 were to be false, it would 
have to be caused by a surge in particle-induced events that were not 
vetoed {\em and} not manifested in the increase of the ``ELECTRON'' factor; 
we are not aware of any examples of such background flaring activities. 
Therefore, we believe that we have detected a sub-hour flare from Mrk 501.
The rare detection can be attributed, to a large extent, to the much 
intensified coverage of the source.

\subsection{Spectral Modeling}

We carried out detailed spectral modeling to investigate, in a more
quantitative manner, spectral variability of Mrk 501. Here, we eliminated 
all observations with effective exposure times less than 200 s, because 
of the lack of statistics. In subsequent analyses, we fixed the hydrogen 
column density at $1.8\times 10^{20}\mbox{ }cm^{-2}$ (Dickey \& Lockman 
1990). We experimented with the following empirical models: power law, 
power law with an exponential rollover, and broken power law, to fit the 
observed X-ray spectra. We found that when the source was relatively weak
(the 2-20 keV flux less than 
$2\times 10^{-10}\mbox{ }ergs\mbox{ }cm^{-2}\mbox{ }s^{-1}$), the power-law
model fits the X-ray spectra well. However, the simple model fails 
statistically (in terms of $\chi^2$ values) at higher fluxes. The data 
taken when the source was brighter (in 1997 and parts of 1998) clearly 
favor the broken power-law model. Formally, the resultant fits are 
satisfactory (with the reduced $\chi^2$ values all around unity). 

Mrk 501 was most active in 1997, with the 2--20 keV flux varying roughly 
between $3$--$9.5 \times 10^{-10}\mbox{ }ergs\mbox{ }cm^{-2}\mbox{ }s^{-1}$. 
The best-fit broken power-law parameters are summarized in Figure 9. The 
first photon index shows an initial decreasing (or spectral hardening) trend 
and then levels off toward high fluxes; so does the second photon index, 
although the trend is less apparent. In general, the SED of Mrk 501 is 
quite flat in the X-ray band during this time period. Spectral variability 
seems to be dominated by changes in the break energy (roughly between 5 and 
16 keV). In the subsequent years (1998--2000), the source became much weaker 
in X-rays, with the flux ranging between 
$0.4$--$4 \times 10^{-10}\mbox{ }ergs\mbox{ }cm^{-2}\mbox{ }s^{-1}$. The
broken power-law model is still required for some of the 1998 observations, 
as already mentioned, and the results are also shown in Fig.~9 for 
comparison. The photon indices show large variations. In some cases, the 
X-ray spectrum is significantly steeper than that observed in 1997. On the 
other hand, the values of the break energy seem comparable to those from 
the 1997 observations. 

At lower fluxes 
($< 2\times 10^{-10}\mbox{ }ergs\mbox{ }cm^{-2}\mbox{ }s^{-1}$), the data 
can no longer effectively constrain the broken power-law model, mainly 
due to the lack of statistics (and also to the fact that the photon indices 
above and below the break do not differ significantly; see Fig.~9). 
Figure 10 shows the results from fits with a simple power law. The trend 
of spectral hardening toward high fluxes is more apparent in this case. 
The photon index goes roughly from 2.6 down to 1.8, clearly indicating a 
shift in the synchrotron SED peak toward higher energies at higher fluxes. 
Also apparent from the figure is spectral hysteresis associated with the 
1998 observations. As shown in Fig.~1, these observations sampled two 
distinct periods: the very end of the decaying phase of the 1997 giant 
flare and the rising phase of a subsequent major flare (of much lower 
magnitude). To examine spectral hysteresis in more detail, we marked 
the data points with different symbols for the two periods in Figures 9 
and 10. For the most part, the points seem to bifurcate, with those for 
each period clustered around one of the branches. However, the phenomenon 
is clearly more complex --- spectral hysteresis also seems to be present 
within each period, perhaps associated with smaller flares that are not 
resolved.

\section{Discussion}

Like Mrk 421, Mrk 501 also produces X-ray flares on a wide range of 
timescales. Unlike in Mrk 421, however, the flares in Mrk 501 do not seem 
to occur nearly as frequently, which appears to be the case on all 
timescales. This would be consistent with the scale invariance of X-ray
flaring phenomenon, if it is common among TeV blazars. Fig.~11 illustrates 
the flaring hierarchy observed in Mrk 501. 
In other words, there might not be any fundamental differences among long 
flares, intermediate flares, or rapid flares, because they could all be 
caused by the same physical process operating on all physical scales. 
Though not fully understood, the flares in blazars are often thought to 
be related to internal shocks in the jet (e.g., Rees 1978; Spada et al. 
2001), or to major ejection events of new components of relativistic 
plasma into the jet (e.g., B\"ottcher et al. 1997; Mastichiadis \& Kirk 
1997). In these scenarios, the flares of different durations might simply 
be the observational manifestation of a hierarchy of inhomogeneities in 
the jet, which are energized to produce flares by the shocks. Alternatively, 
the flares might be associated with magnetic reconnection events in a 
magnetically dominated jet (Lyutikov 2003), perhaps similar to solar 
flares in this regard. The flaring hierarchy might be related to 
reconnection and subsequent avalanche processes. We should note that the 
proton-synchrotron model actually requires a magnetically dominated jet 
to account for the observed TeV emission from blazars (Aharonian 2000; 
M\"{u}cke et al. 2003).

The X-ray spectrum of Mrk 501 evolved significantly during the 1997 and
1998 major flares. It shows a general trend of hardening toward the 
peak of a flare, which is consistent with results from previous 
works on X-ray bright blazars (Giommi et al. 1990), perhaps implying a 
hardening in the spectral distribution of the emitting electrons during 
the major flares. Moreover, there are indications of hysteresis associated 
with the spectral evolution of the source during major flares in 1997 
and 1998. Though common among blazars, spectral hysteresis is still not 
fully understood. Kirk \& Mastichiadis 
(1999) showed, using an internal shock model, that complex hysteresis 
behavior could arise from the interplay of three characteristic timescales 
associated with synchrotron cooling ($\tau_{syn}$), particle acceleration 
($\tau_{acc}$), and intrinsic variability ($\tau_{var}$), respectively.
The model can qualitatively describe the observed spectral hysteresis
associated with rapid flares, but would have difficulty in explaining the
phenomenon associated with long-duration flares in Mrk 501, because 
$\tau_{var}$ almost certainly dominate over $\tau_{syn}$ and $\tau_{acc}$ 
in the latter cases. 

We critically examined the reported sub-hour X-ray flare from Mrk 501 
(Catanese \& Sambruna 2000), in light of intense background flaring
activities during the observation. We found that it was likely associated 
with flaring activities in the flux of soft electrons entering the 
detectors, as opposed to those of Mrk 501. A similar rapid X-ray flare 
was detected in this work, which was also accompanied with a strong 
electron flare. On the other hand, we also found a rapid X-ray flare that 
occurred during a time that seems to be free of any background
flaring activities. The flare lasted only for about 800 s; it might in  
fact be composed of two flares on even shorter timescales. These 
variability timescales pose severe 
constraints on the physical properties of the flaring region. First of
all, the (jet-frame) size of the region is constrained by
$l \lesssim c t_{flare} \delta/(1+z) = 2.4\times 10^{14} \delta_1\mbox{ }cm$,
where $t_{flare}$ is the duration of the flare ($=800$ s), $\delta$ is the
Doppler factor of the jet ($\delta = 10\delta_1$), and $z$ is the redshift 
of Mrk 501 ($= 0.034$). The upper limit is already comparable to the
gravitational radius ($r_g \equiv G M/c^2$) for a $1.3\times 10^9 M_{\odot}$
black hole, which is believed to exist in Mrk 501 (Barth et al. 2003),
if the Doppler factor of the jet is not too much larger than 10 (which 
is probably the case; see discussion below). Since the peak flux of the
flare is a significant fraction of the steady-state flux, the size of
the flaring region is probably comparable to the lateral extent of the jet.
If the jet originates from accretion flows, as is often thought to be the 
case, the result would also represent an upper limit on inner boundary of 
the flows (either in the form of a geometrically thin, optically thick 
cold disk or of a geometrically thick, optically thin hot torus).

Secondly, the observed decaying time of the flare 
sets a firm upper limit on the synchrotron cooling time of the emitting 
electrons. The cooling time is given by Rybicki \& Lightman (1979), 
$\tau_{syn} \approx 6 \pi m_e c/\sigma_T \gamma_p B^2$, where
$m_e$ is the electron rest mass, $\sigma_T$ is the Thomson cross
section, $B$ is the strength of the magnetic field in the region, and 
$\gamma_p$ is the characteristic Lorentz factor of those
electrons that contribute to the bulk of the observed X-ray emission 
(at $E_{p}\sim 10$ keV, where the synchrotron peak of the SED lies).
From $\tau_{syn}/\delta < t_d$, where $t_d$ is the decaying time of the
flare ($\approx 400 s$), we can derive a lower limit on the magnetic field 
strength, $B > 1.4 \delta_1^{-1/2} \gamma_{p,5}^{-1/2} \mbox{ }G$, where 
$\gamma_p = 10^5\gamma_{p,5}$. This is already larger than typical values
of $B$ inferred from fitting the SED with the SSC models (e.g., Kataoka 
et al. 1999; Sambruna et al. 2000; Petry et al. 2000; Krawczynski et al. 
2002; Konopelko et al. 2003). Finally, for synchrotron photons to reach 
X-ray energies (at $\sim E_p$), the Lorentz factor of the emitting 
electrons must be sufficiently high, 
$E_{p}=\delta h\nu_c \equiv (3 e h/4 \pi m_e c) \delta \gamma_{p}^2 B$
(Rybicki \& Lightman 1979). 
Using the derived lower limit on $B$, we can derive an {\em upper} limit
on the Doppler factor, $\delta < 6.4 E_{p,1}^2 (\gamma_{p,5}/3)^{-3}$,
which seems quite low compared to typical values inferred from the SSC 
models.

\acknowledgments

We thank Keith Jahoda and Craig Markwardt for insightful discussions on 
the issue of electron contamination and the referee for his/her very 
detailed and constructive comments. This work was supported in part by 
grants from the US Department of Energy and National Aeronautics and 
Space Administration. We made use of data obtained through the High 
Energy Astrophysics Science Archive Research Center Online Service, 
provided by the NASA/Goddard Space Flight Center.

\clearpage

\begin{figure}
\psfig{figure=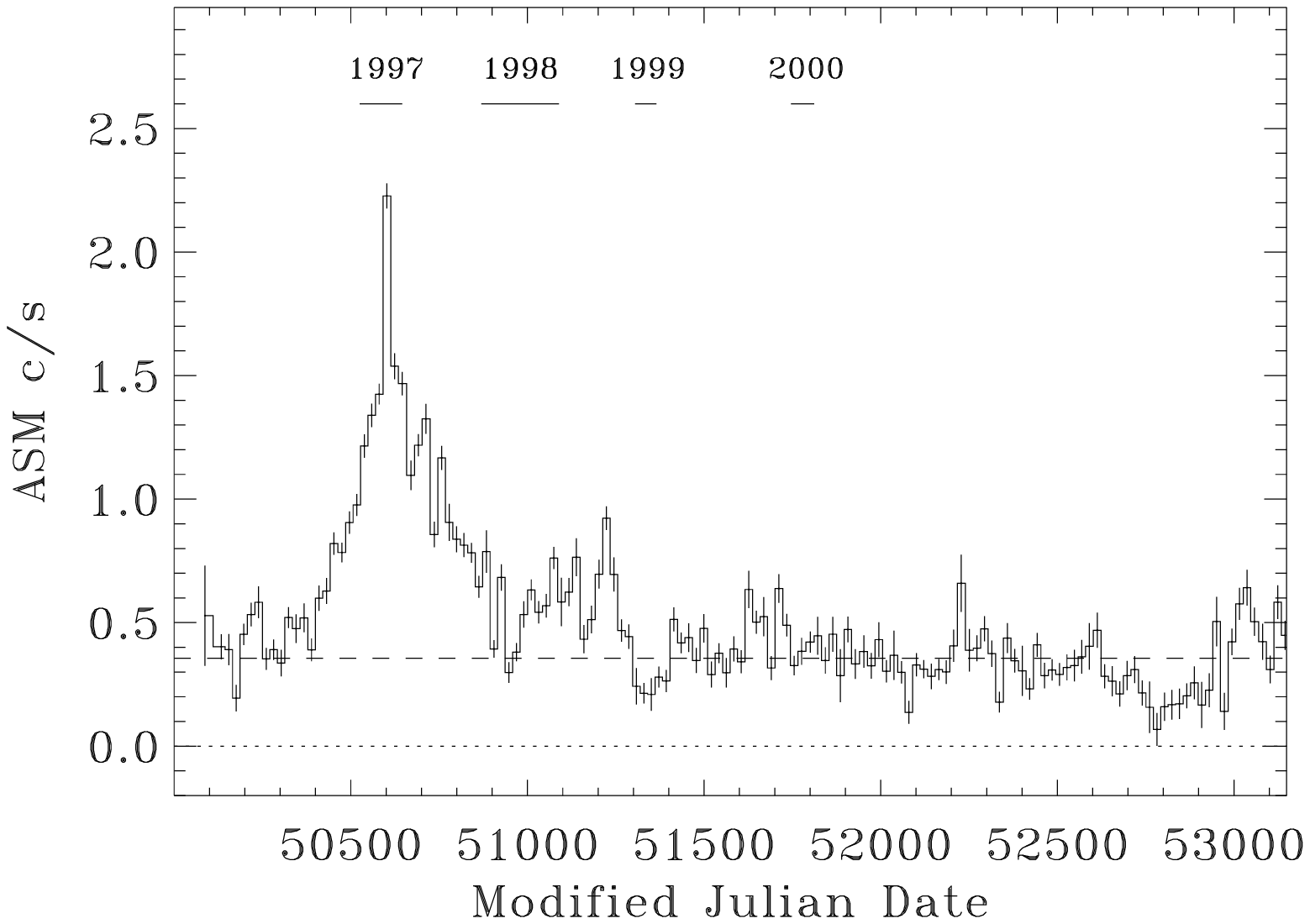,width=5in} \caption{ASM light curve of
Mrk~501. Each data point was derived from a weighted average of
raw count rates over 21 consecutive days (after filtering out the
``bad data''; see text). The error bars shown represent the
standard deviations. The dashed-line roughly shows the average
count rate of the source in the ``low state'' (which is an average
of the rates between MJD 51400--53150). The periods of monitoring
campaigns (with the pointing instruments) are indicated.}
\end{figure}

\begin{figure}
\psfig{figure=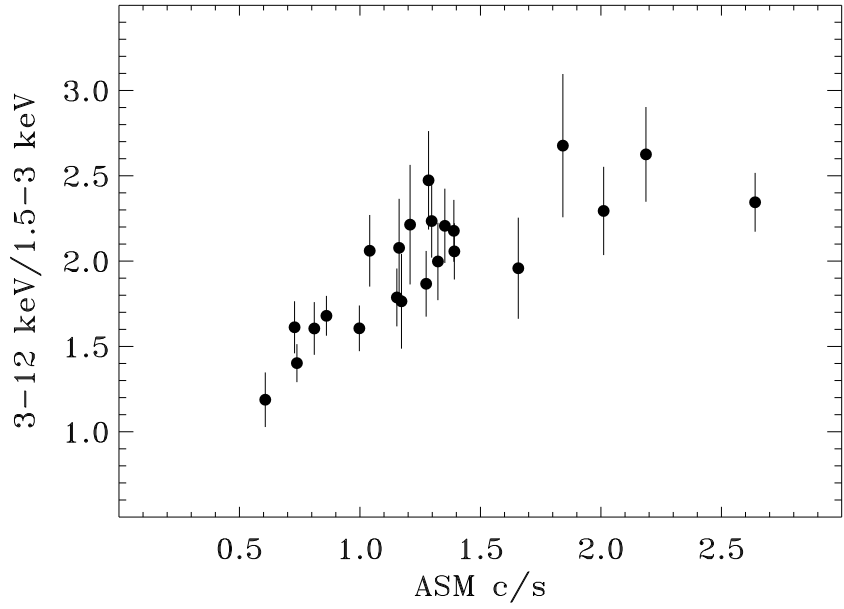,width=5in} \caption{ASM hardness-intensity
diagram for the 1997 giant flare (MJD 50442--50875). Note a
general hardening trend of the X-ray spectrum as the flux
increases. }
\end{figure}

\begin{figure}
\psfig{figure=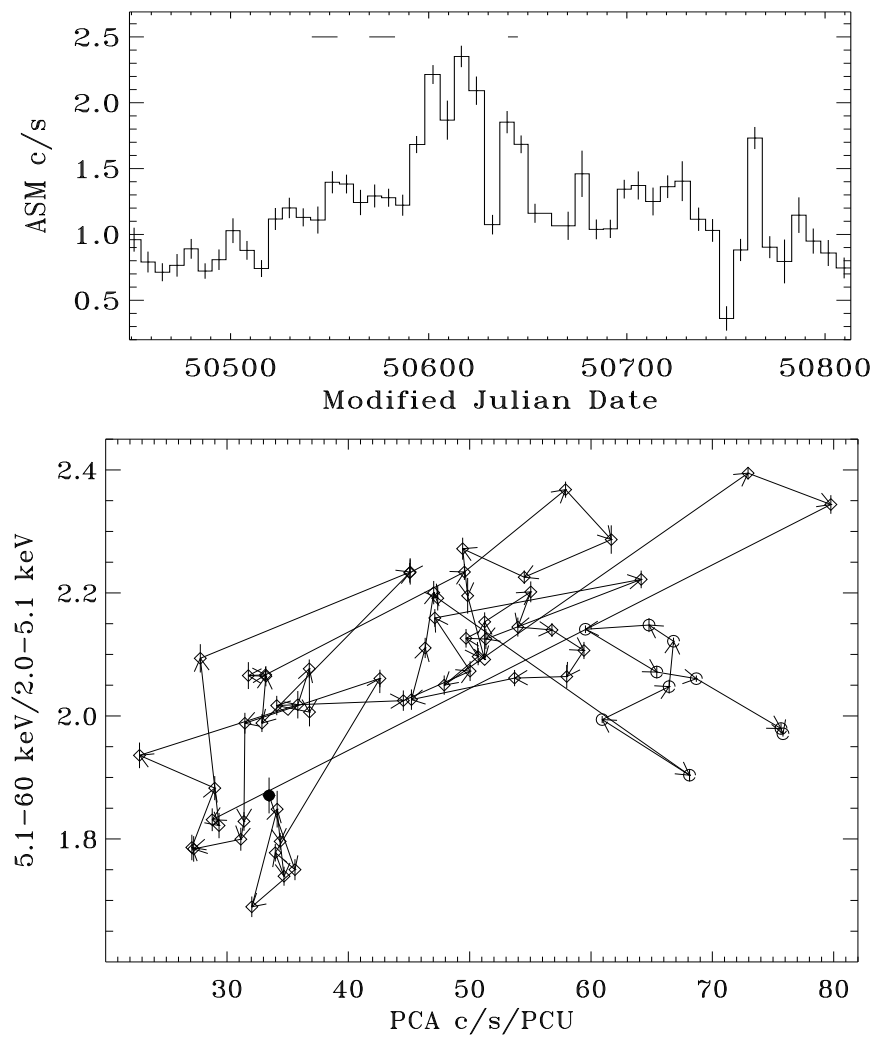,width=5in} \caption{(top) Expanded ASM view
of the 1997 giant flare. The count rates are weekly averaged in
this case. The horizontal lines at the top indicate periods of the
PCA observations. (bottom) PCA hardness-intensity diagram. The
count rates were computed for the 2--60 keV band. Data taken
during the rising portion of the flare are shown in open diamonds
(except for the very first data point, which is indicated by a
filled circle), while data during the decaying phase in open
circles. The direction of the arrows shows the time progression of
the flare. Note the general trend of spectral hardening toward the
peak of the flare. }
\end{figure}

\begin{figure}
\psfig{figure=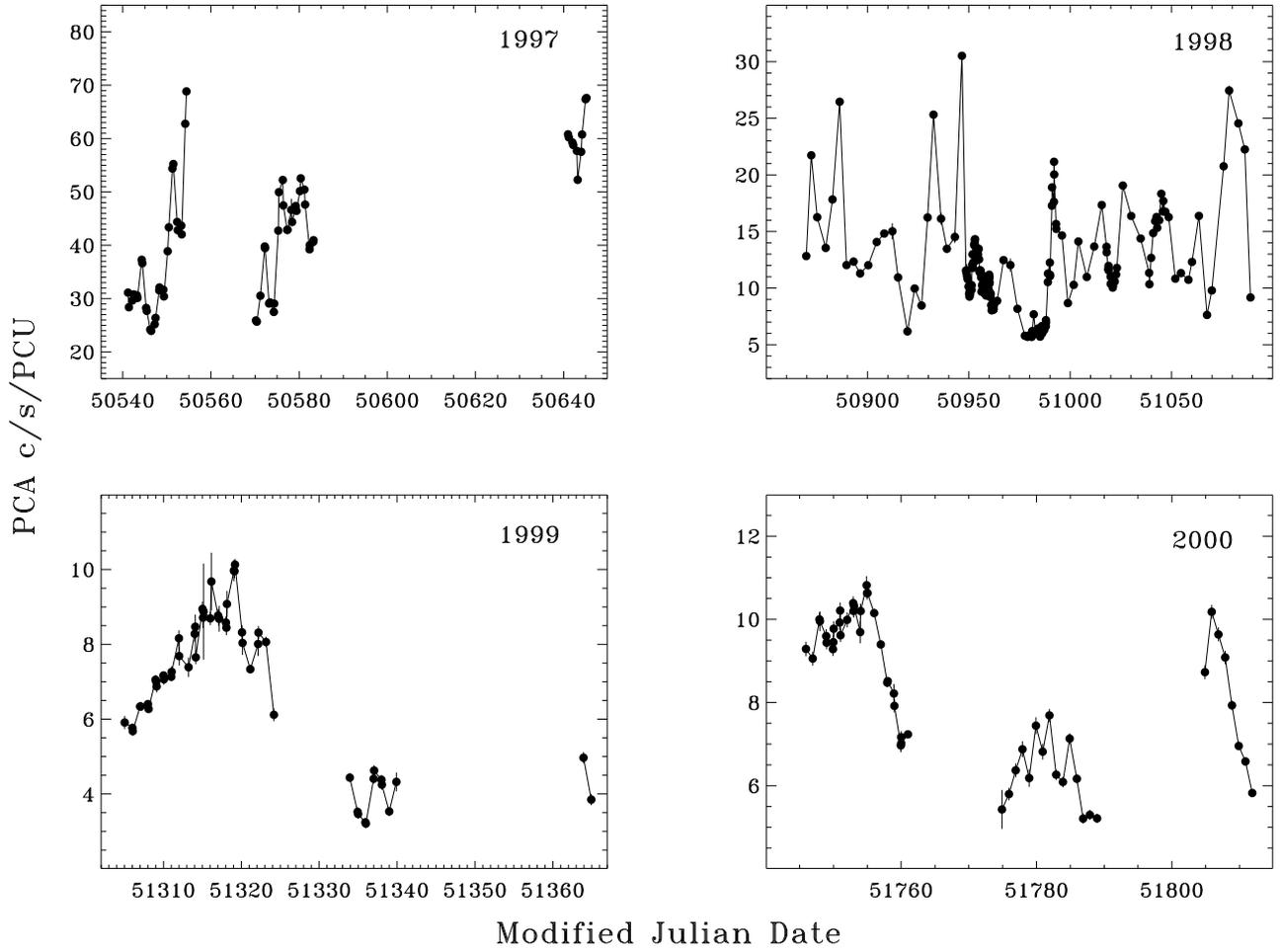,width=5in,angle=90} \caption{PCA light curves
of Mrk~501. Unlike in Fig. 3, the results here were derived from
the PCU 2 data alone. For clarify, they are shown for each year
separately. }
\end{figure}

\begin{figure}
\psfig{figure=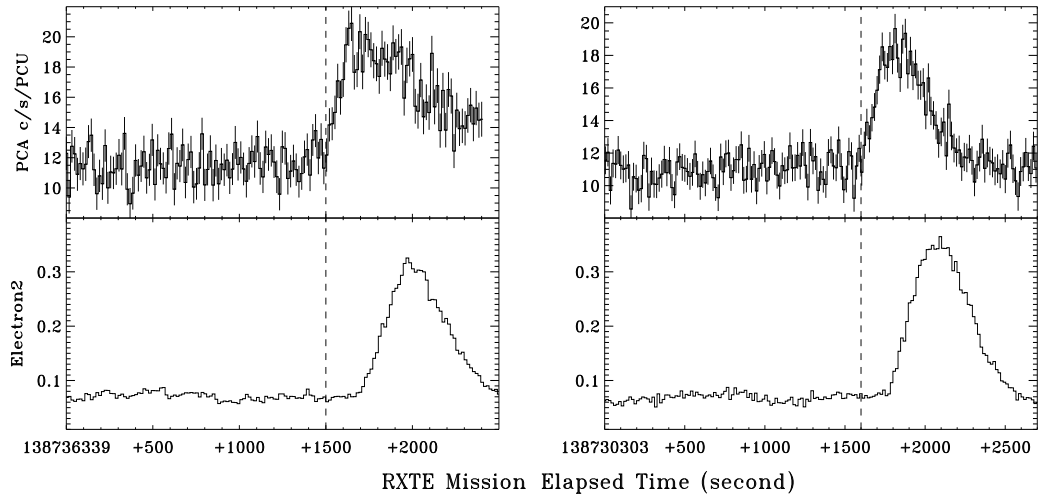,width=6in} \caption{(left) Reported sub-hour
X-ray flare (Catanese \& Sambruna 2000). It is seen only after the
data filtering criterion on the ``ELECTRON2'' parameter is
relaxed. Note the 200-second lead of the X-ray flare with respect
to the electron flare. (right) A similar X-ray flare, with almost
identical 200-second difference between the starting time of the
X-ray and electron flares. For reference, the RXTE Mission Elapsed
Time (MET) is used, which is defined as the number of seconds
since 1994 January 1 00:00:00 (UT). }
\end{figure}

\begin{figure}
\psfig{figure=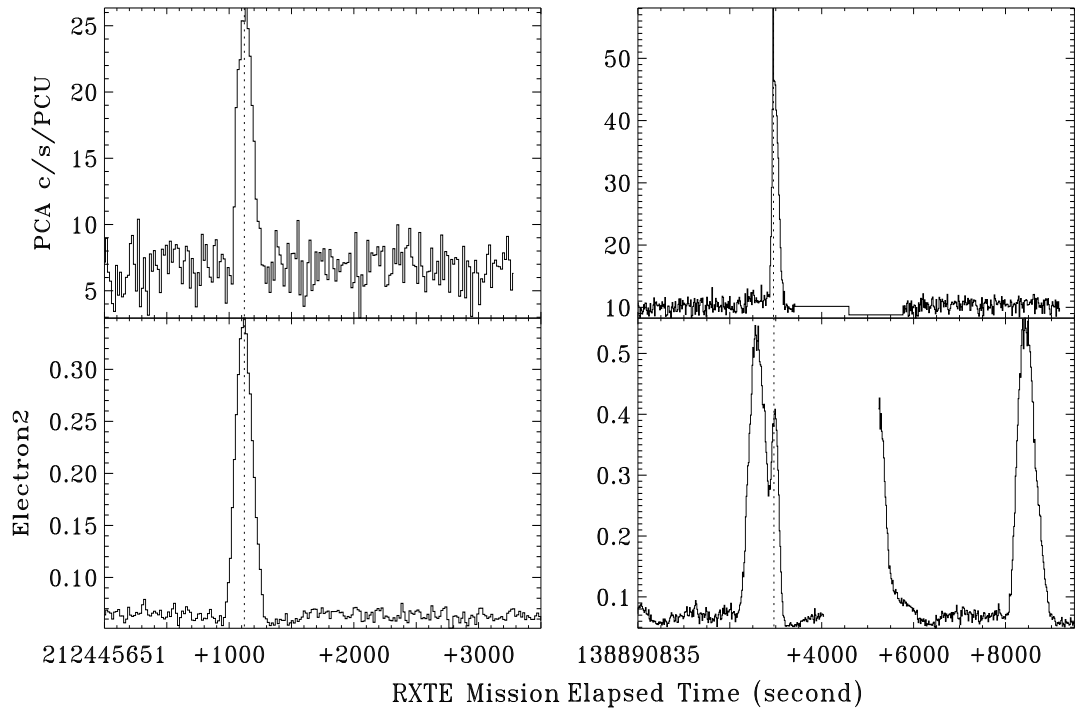,width=6in} \caption{Sample X-ray flares with
associated electron flares. Note that not all electron flares
induce X-ray flares and that an electron flare may have a
complicated multi-peaked profile. }
\end{figure}

\begin{figure}
\psfig{figure=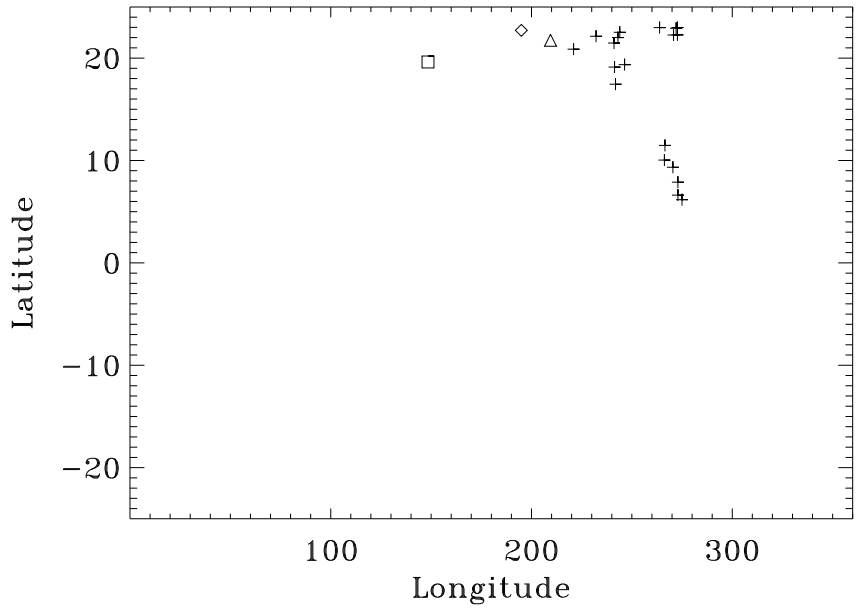,width=5in} \caption{Satellite positions at
the times of the peaks of electron flares. The two flares shown in
Fig.~5 are indicated by different symbols, with the one associated
with the reported X-ray flare in open diamond and the other in
open triangle. Purely for the purpose of comparison, the X-ray
flare shown in Fig.~8 is also indicated here (in open square). It
should be stressed, however, that it is {\em not} accompanied by
any electron flare; in this case, the satellite position was
derived from the starting time of the X-ray flare. }
\end{figure}

\begin{figure}
\psfig{figure=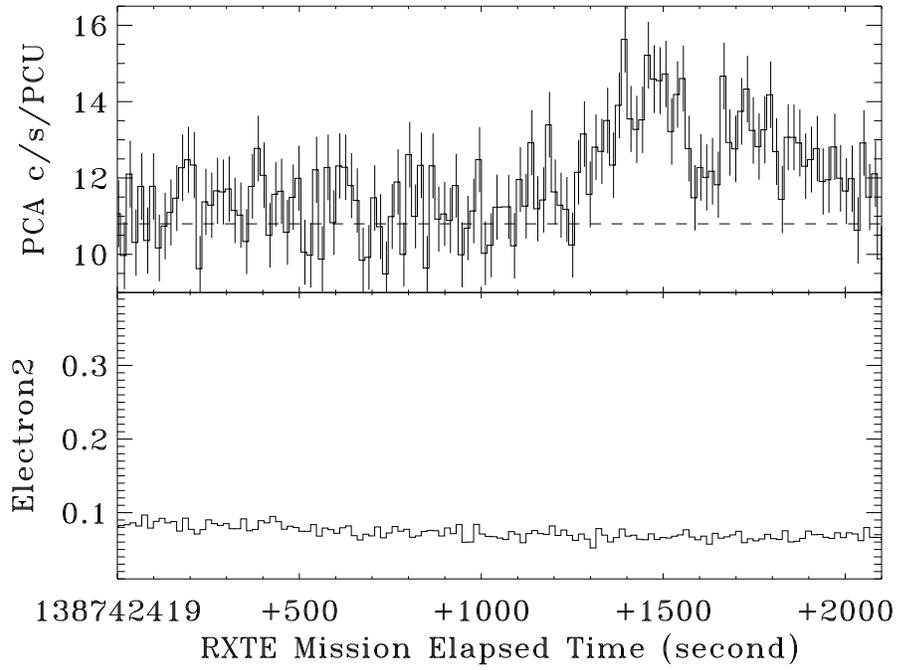,width=5in} \caption{Same as Fig.~5 but for a
real sub-hour X-ray flare from Mrk 501. Note sub-structures in the
profile of the flare. }
\end{figure}

\begin{figure}
\psfig{figure=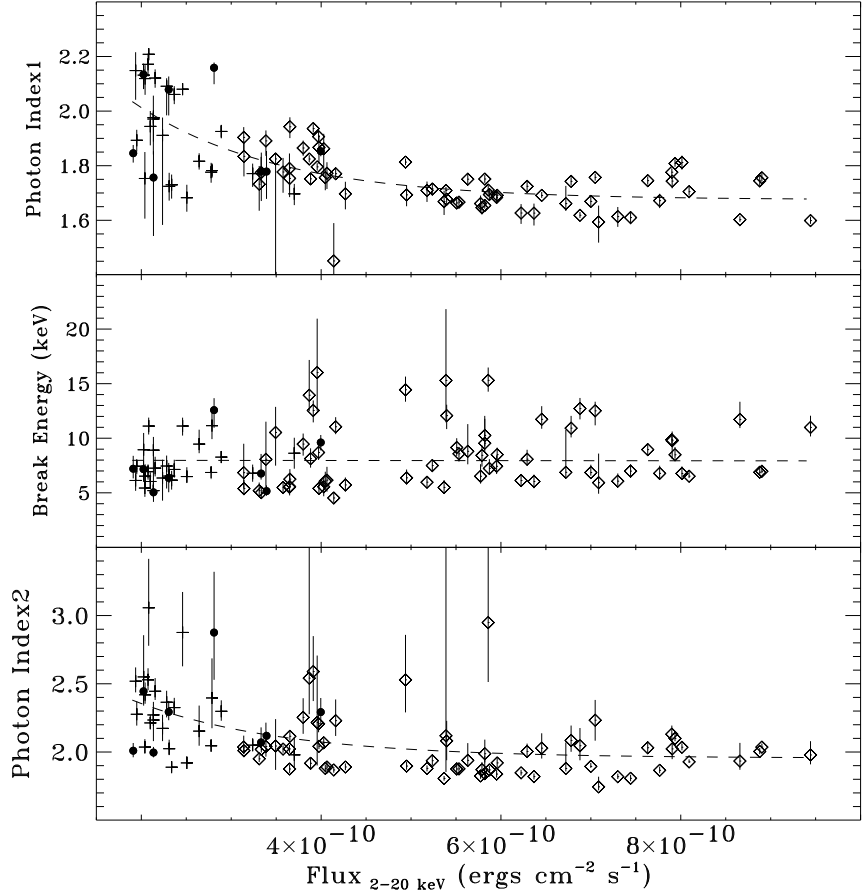,width=5in} \caption{Flux dependence of the
model parameters. The results were obtained with a broken power
law and are shown in diamonds for observations taken in 1997, in
filled circles for those (taken in 1998) near the end of the
decaying phase of the 1997 giant flare, and in crosses for those
during the rising phase of the 1998 flare (with the dividing line
between the two 1998 period drawn somewhat arbitrarily at MJD
50960.6; see Fig.~1). To guide the eye, the dashed lines are drawn
to show best-fits to the data with a model consisting of a
constant plus an exponential function. }
\end{figure}

\begin{figure}
\psfig{figure=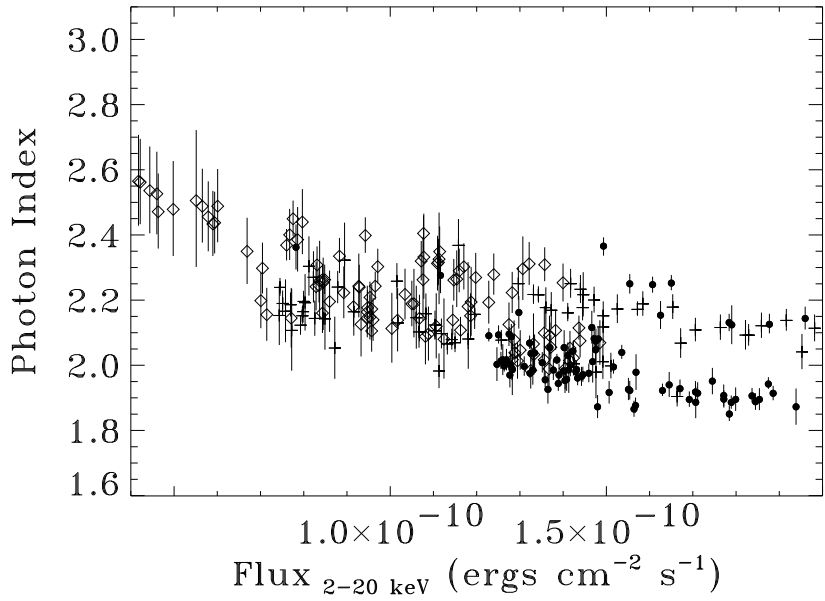,width=5in} \caption{Same as Fig.~9, but a
simple power law was used to fit the data. The results are shown
in diamonds for observations taken in 1999 and 2000, in crosses
for those during the rising phase of the 1998 flare, and in filled
circles for those near the end of the decaying phase of the 1997
flare. }
\end{figure}

\begin{figure}
\psfig{figure=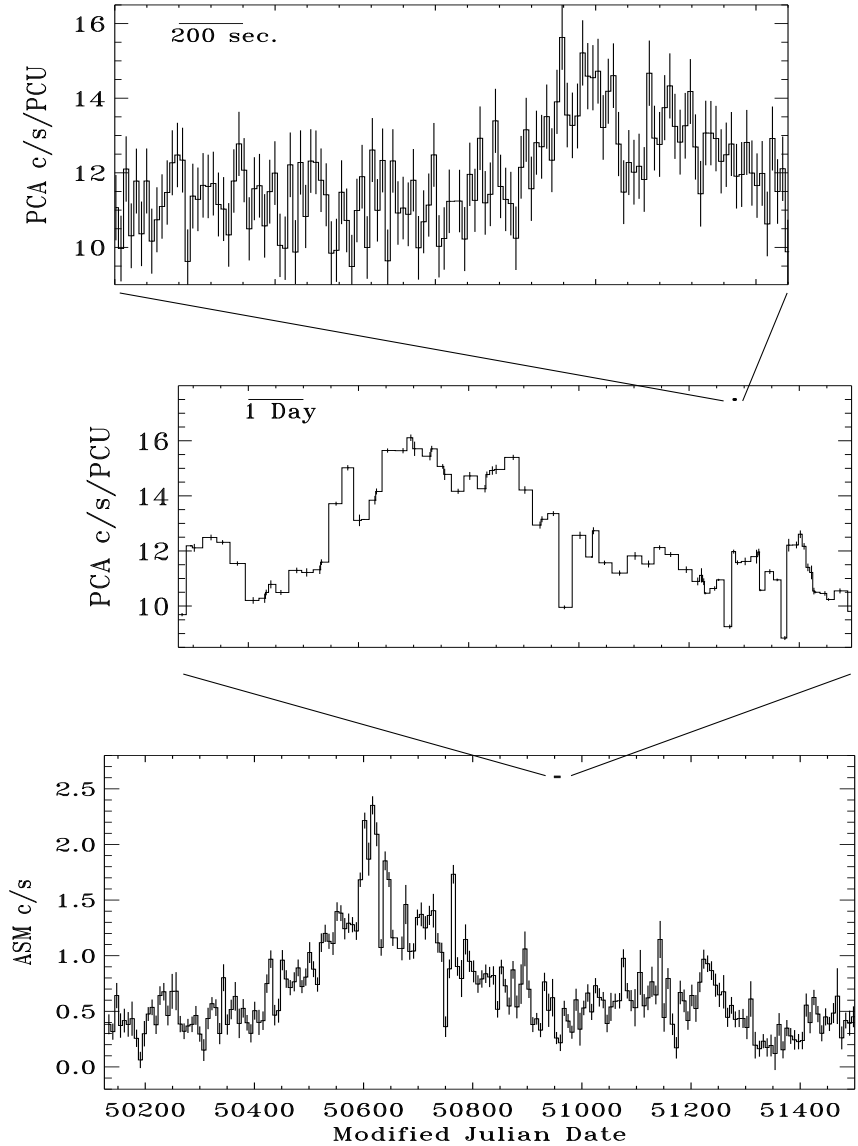,width=5in} \caption{X-ray flares from Mrk
501 on a wide range of timescales. Apparent irregularity in the
width of histogram bins is a manifestation of irregular data
sampling. }
\end{figure}

\end{document}